\documentclass{elsart5p}

\usepackage{graphics}
\usepackage{graphicx}
\usepackage{amssymb}
\begin{document}

\begin{frontmatter}

\title{Energy-scale phenomenology and pairing via resonant spin-charge motion
in FeAs, CuO, heavy-fermion and other exotic superconductors}

\author{Y. J. Uemura}
\ead{tomo@lorentz.phys.columbia.edu}

\address{Department of Physics, Columbia University, 538 West 120th Street, New York, NY 10027, USA}

\corauth[PPP]{Corresponding author. Tel/Fax: +1 212 854 8370}

\begin{abstract}

Muon spin relaxation ($\mu$SR) studies of the ``1111'' and ``122'' FeAs systems have detected 
static magnetism with variably sized ordered moments in their parent compounds.
The phase diagrams of FeAs, CuO, organic BEDT, A$_{3}$C$_{60}$ 
and heavy-fermion systems indicate competition between static magnetism and superconductivity,
associated with first-order phase transitions at quantum phase boundaries.   
In both FeAs and CuO systems, the superfluid density $n_{s}/m^{*}$ at $T \rightarrow 0$ exhibits a 
nearly linear scaling with $T_{c}$.
Analogous to the roton-minimum energy scaling with the 
lambda transition temperature in superfluid $^{4}$He,
clear scaling with $T_{c}$ was also found for 
the energy of the magnetic resonance mode in cuprates, (Ba,K)Fe$_{2}$As$_{2}$, 
CeCoIn$_{5}$ and CeCu$_{2}$Si$_{2}$,
as well as the energy of the superconducting coherence peak observed by angle resolved
photo emission (ARPES) in the cuprates near ($\pi$,0).
Both the superfluid density and the energy of these
pair-non-breaking soft-mode excitations determine the superconducting $T_{c}$ via phase 
fluctuations of condensed bosons.
Combining these observations and common dispersion relations of 
spin and charge collective excitations in the cuprates, 
we propose a resonant spin charge motion/coupling,
``traffic-light resonance,'' expected when the charge energy scale $\epsilon_{F}$ 
becomes comparable to the spin fluctuation energy scale $\hbar\omega_{SF}\sim J$, 
as the process which leads to pair formation in these correlated electron superconductors.   

\end{abstract}


\end{frontmatter}

\section{Introduction}

Discovery of superconductivity in the La(O,F)FeAs system \cite{laofeashosono} and the 
subsequent development of the superconducting RE(O,F)FeAs  
(RE= rare earth La, Nd, Ce, Sm, ...; ``1111'') \cite{oldrefs26} 
and (A,K)Fe$_{2}$As$_{2}$
(A = Ba, Sr, Ca; ``122'') systems 
\cite{oldrefs78,cafe2as2pres,lonzarichpressure} 
have generated renewed excitement and interest in
studies of superconductivity in correlated electron systems.
In this paper, we first review recent muon spin relaxation ($\mu$SR) 
measurements of static magnetic order and superfluid density in
the 1111 and 122 FeAs systems, and compare the results with those 
for high-$T_{c}$ cuprates, organic BEDT, A$_{3}$C$_{60}$ and other exotic 
superconductors.  
We then look into energy scales of the magnetic resonance mode (observed
by inelastic neutron scattering) and of the superconducting coherence peak
(by ARPES), and suggest that both of these are manifestations of 
soft-mode excitations analogous to rotons in superfluid $^{4}$He.
After discussing the roles of the superfluid density and the soft-mode
energy scales in determining $T_{c}$, we will propose a pairing 
mechanism in correlated-electron superconductors based on
a resonant spin-charge motion/coupling, which we term as 
``traffic-light resonance.''  The original ideas of this picture 
were presented in refs. \cite{uemurarotonJPCM,uemurarotonPhysica}.

\section{Magnetic order of the FeAs systems}

\begin{figure*}[t]
\begin{center}
\includegraphics[width=0.95\textwidth]{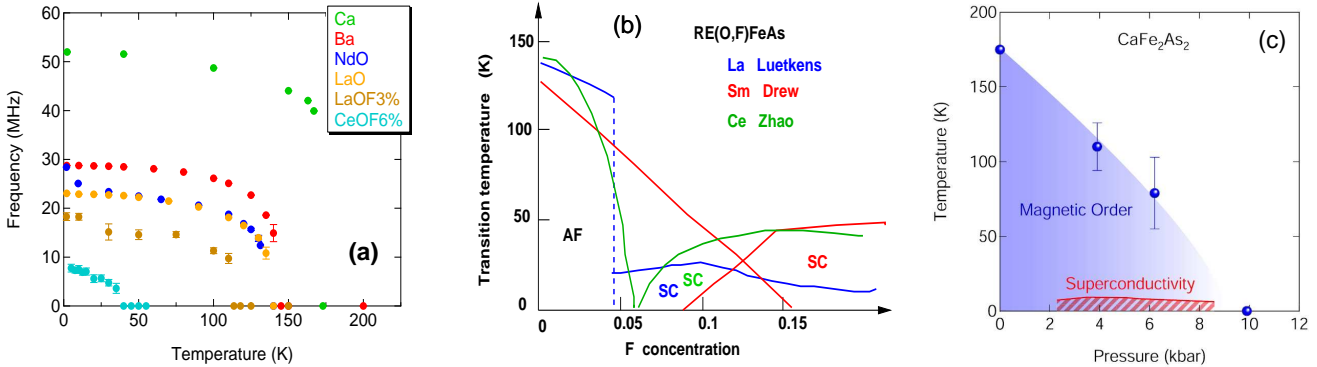}
\end{center}
\caption{(a) Temperature dependence of the muon spin precession 
frequency $\nu$ observed in zero field 
in CaFe$_{2}$As$_{2}$ \cite{gokocondmat}, 
BaFe$_{2}$As$_{2}$, NdOFeAs \cite{aczelcondmat}, LaOFeAs,
La(O$_{0.97}$F$_{0.03}$)FeAs \cite{carlocondmat}, 
and Ca(O$_{0.94}$F$_{0.06}$)FeAs \cite{ceunpublished}.
(b) Phase diagrams of RE(O,F)FeAs systems obtained for 
RE = La by Luetkens {\it et al.\/} \cite{luetkensphase} (blue lines), 
RE = Sm by Drew {\it et al.\/} \cite{drewphase} (red) based on $\mu$SR measurements, 
and RE = Ce by Zhao {\it et al.\/} \cite{daiphase} (green) on neutron measurements 
in ambient pressure.  
(c) Phase diagram of CaFe$_{2}$As$_{2}$ in applied pressure based on $\mu$SR results by
Goko {\it et al.\/}  \cite{gokocondmat} 
for magnetic order and transport results \cite{cafe2as2pres} 
for superconductivity.
} \label{fig1}
\end{figure*}

Static magnetic order in the undoped and lightly-doped compounds 
of the 1111 and 122 FeAs families has been studied by neutron scattering
\cite{daineutron,bafe2as2neutron,cafe2as2goldman},
$\mu$SR \cite{klausscondmat,oldrefs2225,
drewcondmat,carlocondmat,aczelcondmat,gokocondmat} 
and Moessbauer-effect measurements 
\cite{klausscondmat,bafe2as2moess,srfe2as2moessmusr}.
All the undoped systems exhibit long-range collinear antiferromagnetic (AF)
order of Fe moments \cite{daineutron,bafe2as2neutron,cafe2as2goldman}.  
The size of the ordered moment can
be estimated from the hyperfine field in Moessbauer experiments
and $\mu$SR precession frequencies $\nu$ in zero field.
Since the hyperfine coupling constants between Fe moments
and muon spins are nearly equal (difference $<$ 10 \%) for
the 1111 and 122 compounds \cite{aczelcondmat}, 
the $\mu$SR frequency $\nu$ represents the
temperature and system variation of the ordered Fe moment.

Figure 1(a) shows our $\mu$SR results on several non-superconducting 
undoped and lightly-doped 1111 \cite{carlocondmat,aczelcondmat}
and 122 compounds \cite{aczelcondmat,gokocondmat,ceunpublished}.
This figure demonstrates that the size of the ordered Fe moments varies
continuously from $\sim$ 0.1 to 1.0 Bohr magnetons, nearly scaling with 
$T_{N}$.  This feature contrasts clearly with   
cuprate systems with static AF or stripe spin order,
which almost always appears with ordered moment size 
$\sim$ 0.5 Bohr magnetons per Cu at $T \rightarrow 0$.  
The variable moment size in FeAs systems suggests an itinerant character
of 3-d electrons forming multiple bands.  

\begin{figure*}[t]
\begin{center}
\includegraphics[width=0.975\textwidth]{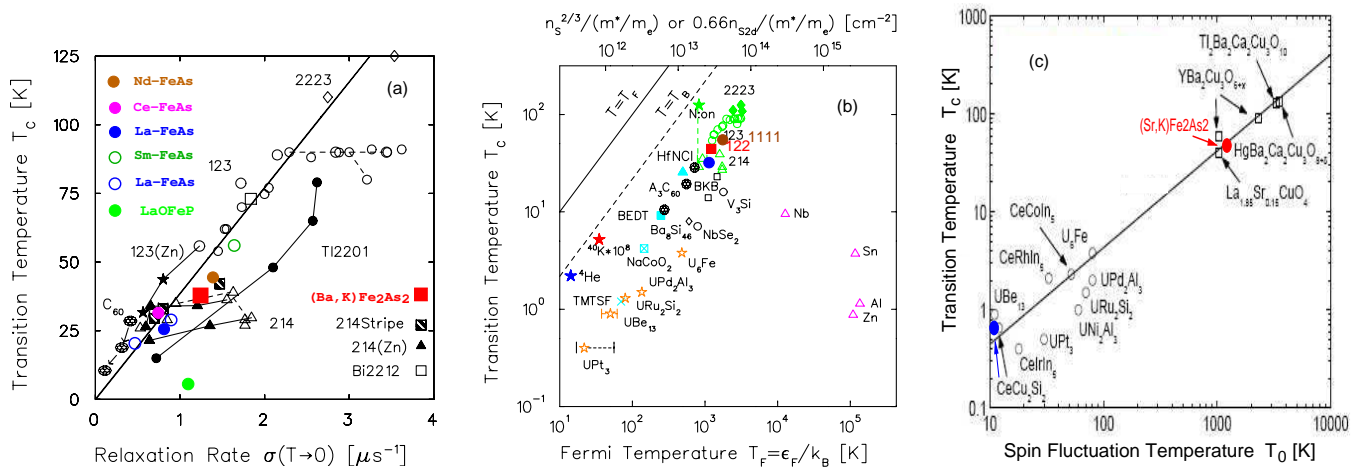}
\end{center}
\caption{(a) Plot of $T_{c}$ versus the muon spin 
relaxation rate $\sigma(T\rightarrow 0)$ observed by 
TF-$\mu$SR measurements on cuprates \cite{uemuraprl89,uemuraprl91,uemuraa3c60nature,lebedtprl,uemurarotonPhysica}, 
1111 FeAs \cite{oldrefs2225,carlocondmat}
(Ba,K)Fe$_{2}$As$_{2}$ \cite{gokocondmat}, LaOFeP \cite{carlocondmat}, 
A$_{3}$C$_{60}$ \cite{uemuraa3c60nature}, 
and various other exotic superconductors
\cite{lebedtprl,uemurarotonPhysica}.  
The relaxation rate $\sigma$ is related to 
the London penetration depth $\lambda$, superconducting carrier density $n_{s}$,
and effective mass $m^{*}$ as $\sigma\propto 1/\lambda^{2} \propto n_{s}/m^{*}$.
(b) Plot of $T_{c}$ versus the effective Fermi
temperature $T_{F}$ derived from the $\mu$SR results on superfluid density
$n_{s}/m^{*}$ for 2-dimensional, and from $n_{s}/m^{*}$ combined with 
the Sommerfeld constant or Pauli susceptibility for 3-d systems \cite{uemuraprl91,uemuraa3c60nature,uemurarotonPhysica}.
$T_{B}$ denotes the Bose-Einstein condensation temperature for a non-interacting
Bose gas of correponding boson density $n_{b} = n_{s}/2$ and mass $m_{b} = 2m^{*}$.
(c) Plot of $T_{c}$ versus spin fluctuation energy scale $T_{0}$,
estimated theoretically based on normal-state transport, susceptibility
and specific-heat results
for cuprates and heavy fermion systems made by Moriya and Ueda 
\cite{moriyaflex}.  Also included are points for 
CeCu$_{2}$Si$_{2}$ (blue solid circle) based on inelastic neutron scattering results
\cite{cecu2si2stassis,cecu2si2uemura} and for
(Sr.K)Fe$_{2}$As$_{2}$ (red solid circle)
based on neutron scattering results of spin waves in SrFe$_{2}$As$_{2}$
\cite{srfe2as2neutron}.
}\label{fig2}
\end{figure*}

Figure 1(b) shows the magnetic and superconducting 
phase diagrams of the 1111 FeAs systems reported for RE=La by
Luetkens {\it et al.\/} \cite{luetkensphase} (blue lines) and RE=Sm by 
Drew {\it et al.\/} \cite{drewphase} (red) both
based on $\mu$SR, and for
RE=Ce by Zhao {\it et al.\/} \cite{daiphase} (green) on neutron measurements.
The evolution from AF to superconducting (SC) 
states occurs with abrupt 
first-order-like change in the La systems, second-order-like change in the 
Ce systems, and associated with a region of coexisting magnetism and superconductivity in 
the Sm systems.
In ZF-$\mu$SR studies
on CeO$_{0.94}$F$_{0.06}$FeAs \cite{ceunpublished} using the same sample  
studied by Zhao et al. \cite{daiphase}, we found a 
short-range static magnetic order developing below T = 40 K as shown in Fig. 1(a).
This makes the RE=Ce phase diagram closer to those of RE=La and Sm.  

Superconductivity in the 122 (A,K)Fe$_{2}$As$_{2}$ (A = Ca,Ba,Sr)
systems can be obtained by hole doping \cite{oldrefs78} 
as well as by application of
hydrostatic pressure to the undoped parent compounds 
\cite{cafe2as2pres,lonzarichpressure}.  
In both cases, our $\mu$SR studies \cite{aczelcondmat,gokocondmat} have 
revealed that superconducting specimens exhibit static magnetic order 
in a partial volume fraction, typically $\sim$50 \%\ of the total volume. 
Figure 1(c) shows the phase diagram for CaFe$_{2}$As$_{2}$ in hydrostatic pressure, 
based on our $\mu$SR results for static
magnetic order \cite{gokocondmat}
and resistivity / susceptibility results \cite{cafe2as2pres,lanlcacondmat} 
for superconductivity.
Magnetic order survives in a partial volume fraction over the superconducting dome,
indicating that magnetism is stronger in the 122 systems than in the 1111 family,
probably due to the more 3-dimensional nature of the 122 systems.

At this moment, there is no information on the length scale of the segregation.
It is also unclear whether superconductivity occurs exclusively in volumes 
without static magnetic order or if it prevails over the entire volume.
Evolution from the AF to SC state is also accosiated with
phase separation in the cuprates near the static stripe order, where $\mu$SR studies
found microscopic segregation with a length scale comparable to the in-plane
coherence length and superconductivity surviving exclusively in volumes without static 
magnetism \cite{saviciLCOPRB,kojimaLESCOPhysica,mohottalaLSCONaturePhys}.  
We consider that this is a likely scenario for the 122 FeAs systems as well.

With increasing pressure and decreasing unit-cell volume,
superconductivity in 
organic (BEDT-TTF)$_{2}$-X \cite{bedtphase} and 
alkali-doped A$_{3}$C$_{60}$ \cite{a3c60prassides} 
appears adjacent to static magnetic order.   Phase diagrams of BEDT, A$_{3}$C$_{60}$, 
cuprates, heavy-fermion, and FeAs systems suggest 
that superconductivity and antiferromagnetism compete
for the ground state.  The evolution is often associated with a first-order transition, as
spatial phase separation was found in the cuprates 
\cite{saviciLCOPRB,kojimaLESCOPhysica,mohottalaLSCONaturePhys},
CeCu$_{2}$Si$_{2}$ \cite{lukeCeCu2Si2PRL}, BEDT \cite{bedtphase} and 122 FeAs \cite{aczelcondmat,gokocondmat} 
while abrupt change was 
found in CeRhIn$_{5}$ \cite{cerhin5}, A$_{3}$C$_{60}$ \cite{a3c60prassides} and RE=La 1111 FeAs
\cite{luetkensphase}.  
Note that superfluid $^{4}$He exhibits similar 
first-order quantum transition from the hcp solid to superfluid state
with decreasing pressure.  The relevance of this phenomenon
shall be discussed in the following sections.

Phase separation between 
volumes with and without static magnetic order has also been detected by
$\mu$SR in the insulating frustrated square lattice J$_{1}$-J$_{2}$ 
system Cu(Cl,Br)La(Nb,Ta)$_{2}$O$_{7}$
at the boundary between collinear AF and spin-gapped states 
\cite{uemuraj1j2prl}.  
This similarity suggests an important role played by spin frustration in 
determining the magnetic states of the FeAs systems \cite{yildirim,siabrahams}.  
The itinerant electron aspect 
and the J$_{1}$-J$_{2}$ frustration should be viewed not necessarily as 
contradictory concepts, but rather as interplaying elements
of the physics in FeAs superconductors.  

\section{Superfluid energy scales}

In type-II superconductors, the London penetration depth $\lambda$
can be determined from the muon spin relaxation rate $\sigma$
defined for the Gaussian relaxation envelope $\exp(-\sigma^{2}t^{2}/2)$
observed in a high transverse external field (TF), as
$$\sigma \propto 1/\lambda^{2} \propto n_{s}/m^{*}\eqno{(1)},$$
where $n_{s}$ is the superconducting carrier density, $m^{*}$ is
the effective mass, and the clean limit is assumed here 
for simplicity \cite{uemuraprl89,reviewrmp,reviewscot}. 
TF-$\mu$SR measurements have been performed 
on several 1111 and 122 FeAs superconductors 
\cite{oldrefs2225,carlocondmat,aczelcondmat,drewcondmat}.
Figure 2(a) shows a plot of $\sigma(T\rightarrow 0)$
versus $T_{c}$.  The results from FeAs systems follow the nearly linear
trend found for cuprates, A$_{3}$C$_{60}$ and organic BEDT systems 
\cite{uemuraprl89,uemuraprl91,uemuraa3c60nature,lebedtprl,uemurarotonPhysica}.
In theories based on BCS condensation, the transition temperature $T_{c}$
is related to the energy scale of attractive interaction represented by the 
gap energy, while one expects only weak and indirect dependence of $T_{c}$ on $n_{s}$.
In contrast, if $T_{c}$ 
is governed by the acquisition of phase coherence
as in Bose-Einstein (BE) condensation of pre-formed pairs,
one expects a direct correlation between $T_{c}$ and $n_{s}$.
The remarkable universality in Fig. 2(a) suggests strong
relevance of the latter case to condensation in all these
exotic superconductors, including FeAs systems.

This point can be further appreciated if we convert the 
horizontal axis to ``effectve Fermi energy'' $\epsilon_{F}$ \cite{uemuraprl91},
which is proportional to $n_{s}/m^{*}$ in 2-dimensional (2-d) systems.
For 3-d systems, $\epsilon_{F}$ can be derived by combining
$n_{s}/m^{*}$ with another parameter, such as the Sommerfeld
constant.  The resulting plot of $T_{c}$ versus $\epsilon_{F}$ is 
shown in Fig. 2(b), where the BE condensation temperature $T_{BE}$ of an 
ideal non-interacting Bose gas is shown by the broken line.
Although their actual $T_{c}$'s are lower than $T_{BE}$ by about a factor of
4-5, cuprates, FeAs, A$_{3}$C$_{60}$, organic BEDT, and
some heavy-fermion systems exhibit the highest ratios $T_{c}/\epsilon_{F}$ of 
$T_{c}$ with respect to the kinetic energy $\epsilon_{F}$ of superconducting carriers.

In Fig. 2(a), we also include a point for LaOFeP \cite{carlocondmat}, 
which does not follow 
the linear trend.  It has been known for many years that the 214 cuprate
systems ``branch off'' from the linear trend \cite{uemuraprl89}, leading the optimally doped
214 LSCO superconductor to have $T_{c}$ about a factor 2 lower than the 
123 YBCO system with the same superfluid density.  These features suggest that
$n_{s}/m^{*}$ is not the sole factor for determination of $T_{c}$.
Closeness to the competing state with static magnetic order is likely the
reason why LSCO and LaOFeP systems have relatively low $T_{c}$'s
with respect to their superfluid densities.  
In Fig. 2(b), we include
a point corresponding to superfluid $^{4}$He.
The superfluid transition of $^{4}$He occurs at T = 2.2 K in ambient
pressure, which is reduced by about 30 \%\ from $T_{BE}$ = 3.2 K for a non-interacting Bose gas
with corresponding boson density and mass.  Understanding the mechanisms for ``reduction'' of 
$T_{c}$ in these cases would help identifying the additional factor(s) which determine $T_{c}$
in correlated electron systems.

Before discussing about competing states, let us also look into a spin energy scale.
Moriya and Ueda \cite{moriyaflex} derived a ``spin fluctuation temperature'' $T_{0}$
from transport, susceptibility and specific-heat measurements 
as the energy scale expected for 
spin fluctuations at the zone boundary.   
As shown in Fig. 2(c), they found a nearly linear correlation between
$T_{c}$ and $T_{0}$.  
Although $T_{0}$ was derived through
a theoretical model, this parameter is close to the actual energy scale
observed in inelastic neutron scattering, as demonstrated by the point
for CeCu$_{2}$Si$_{2}$ (blue solid circle) representing available neutron
results \cite{cecu2si2stassis,cecu2si2uemura}.
In this figure, we also include a point for 
(Sr,K)Fe$_{2}$As$_{2}$ (red solid circle) based on its $T_{c}$ and spin wave dispersions 
measured in SrFe$_{2}$As$_{2}$ \cite{srfe2as2neutron}.
Comparison of Figs. 2(b) and 2(c) reveals a remarkable correspondence of charge ($T_{F}$)
and spin ($T_{0}$) energy scales in many exotic superconductors.  This feature shall be
discussed in sections 6 and 7.

\section{Magnetic resonance mode and He rotons: 
soft-mode excitations towards competing states}

In the case of superfluid $^{4}$He, collective excitations of rotons provide a channel
for thermal depletion of condensed bosons.  Thanks to the large phase-space factor
for substantial momentum transfer, the roton-minimum energy $\hbar\omega_{roton}$ 
plays a dominant role in determining superfluid
$T_{c}$, as was noticed by Laudau and Ruvalds \cite{ruvaldsprl}.  
The experimental confirmation of 
this concept can be obtained by plotting early neutron results \cite{neutronroton} of 
$\hbar\omega_{roton}$ versus the superfluid
lambda temperature $T_{c}$ in ambient and applied pressure.
The nearly linear relationship shown in Fig. 3 verifies that the roton energy determines
$T_{c}$ of the superfluid state.  Rotons are soft-mode phonons related to solid
hcp He, whose energy reflects the ``closeness'' of the superfluid state to 
the adjacent and competing solid He state.  Here we find a good example of 
how the competing state influences superconducting or superfluid transitions, by
providing thermodynamically excitable soft collective modes.  
This process does not involve ``pair-breaking,'' but rather  
puts a condensed boson into a state with different phase.  Soft-mode excitation
is thus a process of phase fluctuation, distinct from the Kosterlitz-Thouless process
due to low dimensionality \cite{kttheory}.

\begin{figure}[t] 
\begin{center}
\includegraphics[width=0.45\textwidth]{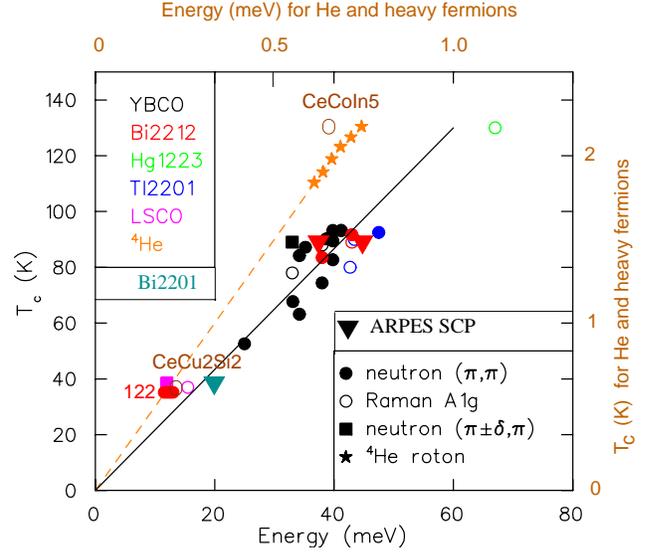}
\end{center}
\caption{Plot of $T_{c}$ versus collective mode energy
$\hbar\omega$ for rotons in superfluid $^{4}$He in ambient and applied pressure 
(orange stars) \cite{neutronroton}, magnetic resonance mode (MRM) in various cuprates (solid circles) \cite{neutronmrmcuprates}, 
spin-gap energies in LSCO and YBCO (solid squares) 
\cite{christensenprl,tranquadalbco},
MRM in (Ba,K)Fe$_{2}$As$_{2}$ (red flat solid circle) 
\cite{mrm122}, CeCoIn$_{5}$ (brown tall open circle) \cite{cecoin5mrm}, 
CeCu$_{2}$Si$_{2}$ (brown tall open circle) \cite{cecu2si2mrm}, 
Raman A$_{1g}$ mode energy in cuprates (open circles) \cite{ramana1g}, 
ARPES SCP peak energies in cuprates 
(solid reverse triangles) \cite{arpesrmp,arpes2008}.
The horizontal and vertical axes set for He and heavy fermions (brown axis label) has the 
same aspect ratio as that for other systems (black label), to facilitate
comparisons of the slope $T_{c}/\hbar\omega$.}\label{fig3}
\end{figure}

The magnetic resonance mode (MRM), observed by inelastic neutron 
scattering, is a likely candidate for
an analoguous soft mode in correlated-electron superconductors.
Figure 3 demonstrates that the mode energy $\hbar\omega_{MR}$ of 
the MRM of various cuprates \cite{neutronmrmcuprates} 
scales with their
$T_{c}$, and the ratios $T_{c}/\hbar\omega_{MR}$ (slope in Fig. 3) are comparable to that of 
rotons.  The MRM is a spin soft mode related to the stripe spin-charge ordered state in the cuprates.
Recent neutron studies revealed an ``hour-glass-shape'' dispersion relation of this mode 
\cite{christensenprl,tranquadalbco}.
For consideration of thermodynamic effects, one should use the low energy peak of 
populated states on
the hour-glass dispersion, which is often denoted as the ``spin gap energy.''  The spin gap energy, shown
by solid square symbols in Fig. 3, exhibits even better agreement 
with the slope $T_{c}/\hbar\omega_{roton}$ of rotons in $^{4}$He.

\begin{figure*}[t]
\begin{center}
\includegraphics[width=0.90\textwidth]{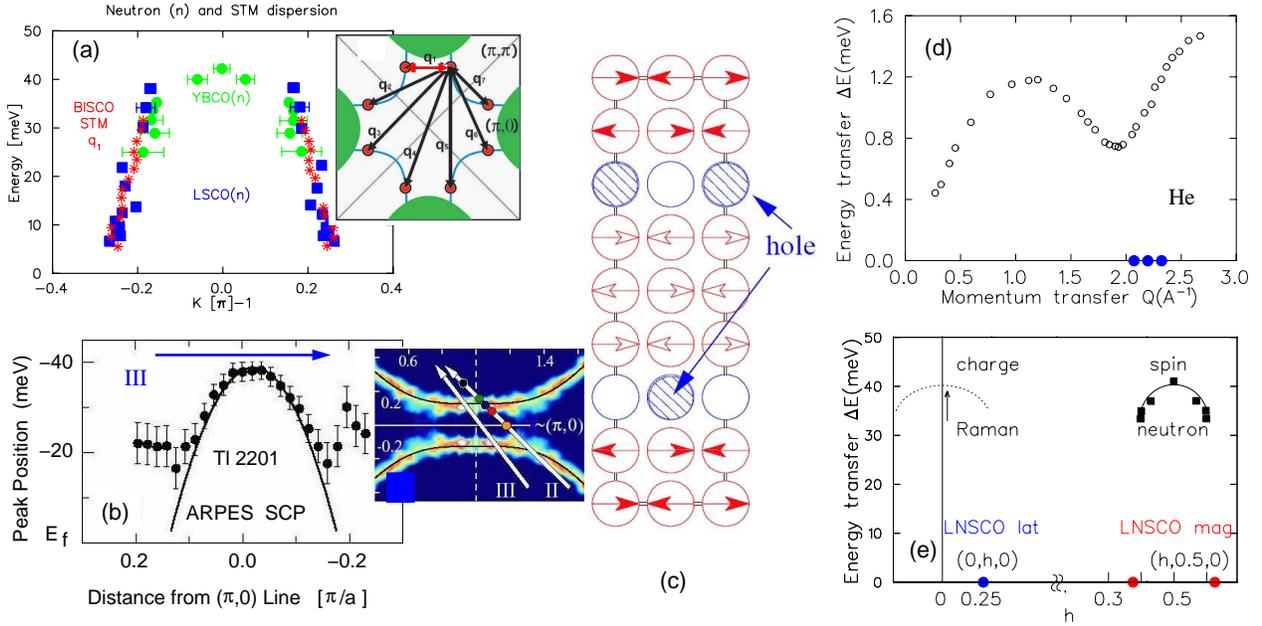}
\end{center}
\caption{Dispersion relation of (a) neutron magnetic resonance
excitations in YBCO and LSCO \cite{christensenprl} cuprates, 
compared with that
of the $q_{1}$ charge mode observed in STM \cite{stmq1} in Bi2212, and (b) 
ARPES SCP peak energy in overdoped Tl2201 near ($\pi$,0) \cite{arpestl2201}.
(c) Spin-charge correlation pattern for stripe correlations in cuprates 
\cite{lnscostripe}.
(d) Dispersion relation of rotons in superfluid $^{4}$He
\cite{rotondispersion}, with solid circles
showing Bragg points of hcp solid He \cite{solidheneutron}.  
(e) Proposed dispersion relation for
twin spin and charge soft modes in cuprates 
\cite{uemurarotonJPCM,uemurarotonPhysica} with blue and red solid circles
denoting Bragg peaks due, respectively, to charge and spin correlations 
\cite{lnscostripe}.}\label{fig4}
\end{figure*}

Very recently, the MRM was also observed in (Ba,K)Fe$_{2}$As$_{2}$ 
\cite{mrm122}, 
CeCoIn$_{5}$ \cite{cecoin5mrm},
and CeCu$_{2}$Si$_{2}$ \cite{cecu2si2mrm}.  Amazingly, the points for 
these systems closely follow the 
roton relationship in Fig. 3.  In general, the excitation energy of $\sim$4-5$k_{B}T_{c}$ 
in Fig. 3 can be
expected for the roton-like pair-non-breaking excitations as well as 
for pair-breaking single-particle excitations
across the d-wave energy gap.  For the former, we shall expect responses sharply 
localized in momentum space 
corresponding to the Bragg peak position of the competing magnetic state.  
In contrast, pair-breaking
single-particle excitations must have a broad intensity profile in momentum space.  
The MRM's observed in all these
exotic superconductors exhibit very sharp profiles in momentum space, supporting our 
interpretation of collective soft modes associated with competing states.
Note that rotons and MRM's are fluctuations characteristic to first-order quantum phase transitions. 

\section{Spin-charge twin soft mode and common
dispersion relation}

As shown in Fig. 3, the energy scale of the  
MRM spin response is also followed by energies of charge excitations
in cuprates, i.e., the A1$_{g}$ response of Raman measurements 
\cite{ramana1g}
at the zone center
and the superconducting coherence peak (SCP) energy of 
angle resolved photo emission (ARPES) measurements observed 
near ($\pi$,0) \cite{arpesrmp}.  Additionally, Fig. 4e of ref. 
\cite{arpes2008} clearly demonstrates that 
the ARPES SCP ``peak'' energy of various cuprate systems corresponds to the 
neutron MRM energy.
Furthermore, the energy-momentum dispersion of the spin and charge
soft mode responses are nearly identical to each other, as shown
by Fig. 4(a) which compares the MRM hour-glass dispersion \cite{christensenprl}
with 
that of the low-momentum ``$q_{1}$'' charge mode response in scanning tunneling microscope
(STM) measurements 
\cite{stmq1}, and by Fig. 4(b) which shows the energy input for
the SCP response of ARPES measurements in the Tl2201 system near ($\pi$,0) 
\cite{arpestl2201}.

Figure 4(c) shows the spin-charge modulation pattern of stripe correlations 
in the cuprates.  When the magnetic state wins against the competing 
superconducting state, this pattern becomes static, accompanied by spin and 
charge Bragg peaks \cite{lnscostripe}
shown by the solid circles in Fig. 4(e).  
When the superconducting state wins, this 
pattern becomes dynamic and short-ranged soft-mode correlations.
Even in such a dynamic/inelastic situation, one would expect strong spin and
charge coupling.  In other words, by paying the energy cost of mode energies,
inelastic spin and charge probes detect responses from temporary and
short-range stripe patterns.  In this situation, it is natural to expect an identical
energy cost for the corresponding spin and charge excitations.   

The MRM represents collective spin-wave excitations from this short-ranged
spin-charge pattern.  The excitation energy $\sim$10-50 meV of the MRM, much lower
than the exchange energy $J\sim$100 - 200 meV in the cuprates, indicates that this
process does not involve ``pair-breaking'' which costs energy $J$.  
Similarly, the ARPES SCP peak
is produced when a charge is knocked out from the temporary stripe pattern, where
no pair-breaking energy $J$ is required thanks to local spin frustration adjacent
to charges.  Thus, we expect the existence of twin
spin and charge excitations, the former near ($\pi,\pi$) and the latter near
the zone center, as illustrated in Fig. 4(e) and compared with the dispersion of
rotons in superfluid He \cite{rotondispersion} and Bragg peaks of hcp solid He 
\cite{solidheneutron} in Fig. 4(d).  This feature explains why the 
Raman energy corresponds one-to-one with the neutron MRM energy, instead of the 
commonly-observed ``two-magnon Raman'' response appearing with twice the spin
excitation energy.  

\begin{figure*}[t]
\begin{center}
\includegraphics[width=0.90\textwidth]{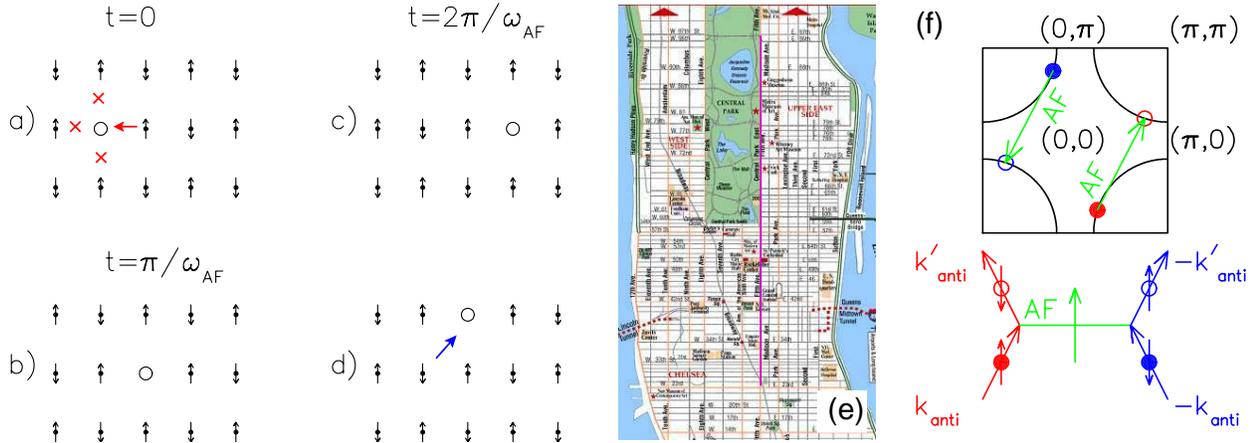}
\end{center}
\caption{Illustration of hole motion in cuprates having
antiferomagnetic (AF) spin correlations.  (a) shows three frustrated bonds
created when the AF pattern is static.  (b) and (c) depict charge
motion occurring sequentially with spin fluctuations. (d) demonstrates no
frustration for nodal charges moving diagonally on the same spin sublattice.
(e) A map of Manhattan, with Fifth Avenue (purple line) where traffic
light alternations are coordinated with the velocity of moving cars.
(f) shows pair formation among antinodal charges via 
scattering of spin fluctuations 
\cite{uemurarotonJPCM,uemurarotonPhysica}.}\label{fig5}
\end{figure*}

These inelastic collective soft-mode excitations in cuprates
are characterized by an unusually strong spin-charge coupling and sharply selective momentum
transfers.  The strong spin-charge coupling also explains why the
excitation of the spin branch in MRM contributes to the depletion of condensed bosons,
thus influencing the superconducting $T_{c}$ analogously to the case of rotons.
The closer the superconducting state is to the competing magnetic state,
the lower is the energy of the soft-resonance mode, consequently 
resulting in the lower $T_{c}$.
Although it is yet to be verified experimentally, 
one may well expect a similar situation in other strongly-correlated superconductors 
existing adjacent to competing magnetic states. 

\section{Spin-charge coupling and pairing via ``traffic-light resonance''}

Now we shall consider microscopic processes which lead to strong spin-charge coupling,
absence of superconductivity in regions with static magnetic order, and strong directional dependence
of the charge coherence as found in the difference 
between nodal and antinodal carriers in the cuprates.
Figure 5(a) illustrates charge motion in cuprates in the environment of antiferromagnetic spin 
correlations.  When the spin pattern is static, the hopping motion of a hole in the antinodal 
direction (horizontal or vertical direction in Fig. 5(a)) immediately results in an energy cost
due to spin frustration.  This frustration, however, can be avoided when the hopping occurs 
sequentially with spin fluctuations.  Specifically,
when the time scale of the charge motion is comparable to half the period of 
spin fluctuation (Figs. 5(b) and (c)), a hole can move without the energy cost
of spin frustration.  This motion can occur successively and coherently, as a resonant charge motion,
if, and only if, the charge energy $\epsilon_{F}$ is comparable to that of spin fluctuations
$\hbar\omega_{SF}$.  For nodal charges propagating in the diagonal direction,
this mechanism is irrelevant, since such holes are moving on the same spin sublattice
(see Fig. 5(d)).

The propagation of antinodal charges in this process, assisted by and resonating with 
spin fluctuations, reminds the author of a car or taxi going
through synchronized traffic lights.  On several avenues in Manhattan
alternation of traffic lights are matched with the velocity of cars.
For example, on Fifth Avenue (Fig. 5(e) purple line) or Amsterdam Avenue, 
a car can travel the entire
route without being stopped
more than a few times.  Here the ``red'' traffic light corresponds to the 
energy cost due to spin frustration or spin selction rules.  Thus, we shall
term this spin-assisted charge motion as ``traffic light resonance.''

Analogous to the charge motion in a usual metal which causes lattice deformation 
via electron-phonon coupling, motion of antinodal charges has to be accompanied by
the cooperating spin fluctuations.  This spin fluctuation can assist motion of another 
charge with opposite momentum, resulting in an attractive interaction via
scattering process as illustrated in Fig. 5(f).  Since the charge energy
scale $\epsilon_{F}$ is comparable to the energy of mediating bosons (spin fluctuations)
$\hbar\omega_{SF}$, this spin-mediated pairing is not retarded, and one can expect
a large attractive interaction.  This mechanism, occuring selectively for antinodal
carriers, is a good candidate for the origin of pseudo-gap formation.
Static magnetic order, as frozen traffic lights, inhibits charge motion and pairing. 

This picture for cuprates, with charges dynamically avoiding spin obstacles
in cooperation with spin fluctuations, may be generalized to most other
correlated electron systems, such as heavy-fermion systems, 
where the ``correlation'' originates from spin frustration
and/or spin selection rules.  In this picture, the ``glue'' includes
all the antiferromagnetic spin fluctuations, with energies extending from the spin gap
energy to $J$.  Carriers near the Fermi surface couple with 
high-energy antiferromagnetic spin fluctuations as $\epsilon_{F}\sim J$.  
All the ``paramagnon''-like fluctuations, supported
by the short-ranged temporal spin-charge correlations, help superconducting 
coupling as their glues.  Thus, the ``competing magnetic states'' cooperate
with superconducting states, supplying the system with useful inelastic fluctuations.
Overdoping with $\epsilon_{F} > J$, however, no longer helps superconductivity, since
excessively high-energy charges cannot find their spin partners.

\section{Discussions: reconciling conflicting approaches}

For many years, theories for cuprates have been discussed in two different
approaches.  Starting from the parent Mott insulator and underdoped region,
which we might call the ``left wing'' of the phase diagram, one finds relevance to 
Mott insulator, strong coupling,
pre-formed pairs, pseudo-gap, and BE condensation.  In these concepts, $T_{c}$ is
related to the particle density, $n_{s}/m^{*}$, $T_{BE}$ and $\epsilon_{F}$, since
pairs are already formed well above the condensation temperature.
Most of the arguments in the present paper belong to this ``left-wing'' approach.  
The other (``traditional'' or ``right wing'') approach starts from the overdoped region, 
involving the concepts of Fermi liquid,
BCS condensation, weak coupling, etc., and treating spin fluctuations as perturbations. In the
``right-wing''
approach, $T_{c}$ is related to the energy scales of ``glues'' and attractive
interactions, since $T_{c}$ implies the pair-formation temperature.
  
Perhaps a representative work of this latter approach is the spin-fluctuation 
theory and FLEX approximation by Moriya and Ueda \cite{moriyaflex}.  
Identifying spin fluctuations as the mediating bosons in the ``right-wing'' 
approach, they generated the plot shown in Fig. 2(c). 
This figure should be compared and contrasted with 
the $T_{c}$ versus $\epsilon_{F}$ plot of Fig. 2(b) which represents the ``left-wing''
phenomenology.  We notice that the energy scale $k_{B}T_{0} \sim \hbar\omega_{SF}$
is very close to $\epsilon_{F}$ estimated from the superfluid density in cuprates and 
several other correlated-electron superconductors which have relatively high ratios of 
$T_{c}/\epsilon_{F}$.   
For pairing mechanisms based on the ``traffic light resonance,'' one requires
$\epsilon_{F}$ to be comparable to $\hbar\omega_{SF} \sim k_{B}T_{0}$.  Therefore,
this mechanism naturally leads to ``reconciliation'' of the left- and right-wing
approaches and provides reasoning for why Figs. 2(b) and 2(c) look alike for
some cuprate and heavy fermion systems.  
On the other hand, it should also be noted that models 
developed from the ``right wing'' approach do not explain correlations between
$T_{c}$ and $n_{s}/m^{*}$ in the underdoped region shown in Fig. 2(a).

Another apparent duality can be found in the role of competing magnetic 
states on superconductivity.  The soft-mode argument identifies closeness 
to the magnetic state as a destructive factor which reduces the superconducting $T_{c}$.  
The ``traffic light resonance,'' however, requires an inelastic excitation associated with 
dynamical correlations originating from the competing magnetic
state.  Charge doping will help increase $T_{c}$ as shown in Figs. 2(a) and (b),
while excessive charge doping would destroy/weaken the magnetic correlations required
for pairing.  This duality should be the origin of formation of the superconducting dome
in phase diagrams of cuprates and several other systems.  

One remaining challenge for our phenomenology is to understand the origin of 
an apparent upper limit $T_{c}/T_{BE} \sim 1/4$ suggested by Fig. 2(b).
Better understanding of the interplays between spins and charges, the left- and right-wing
approaches, superconducting and magnetic states, and BE-BCS crossover would hopefully 
explain this trend.

\section{Acknowledgments}

The author would like to thank collaborations with 
G.M. Luke, T. Goko, E. Saitovitch, H. Kageyama and many other scientists
coauthoring refs. [14-16,19,22,25,28,29,31,34,37-39,41,42] in 
experiments on various exotic superconductors and magnets, and H. Aoki and K. Ueda
for discussions on the Moriya-Ueda theory.
This work is supported by the NSF MWN-CIAM program
DMR 05-02706, 08-06846.

\end{document}